\documentclass{desyproc}

\begin{document}
\title{Searching for Axion-like Particles with Active Galactic Nuclei}

\author{{\slshape Clare Burrage$^1$, Anne-Christine Davis$^2$, Douglas J. Shaw$^3$  }\\[1ex]
$^1$Theory Group, Deutsches Elektronen-Synchrotron DESY, 22607
Hamburg, Germany\\
$^2$Department of Applied Mathematics and Theoretical Physics,
Centre for Mathematical Sciences, Cambridge CB3 0WA, United Kingdom\\ 
$^3$Queen Mary University of London, Astronomy Unit,
School of Mathematical Sciences, Mile End Road, London E1 4NS, United Kingdom }

\contribID{lindner\_axel}

\desyproc{DESY-PROC-2009-05}
\acronym{Patras 2009} 
\doi  

\maketitle

\begin{abstract}
Strong mixing between photons and axion-like particles in the magnetic
fields of clusters of
galaxies  induces a scatter in the observed luminosities of compact
sources in the cluster. 
This is used to construct a new test for
axion-like particles; applied to observations of active galactic
nuclei it is strongly suggestive of the existence of a light
axion-like particle.\\
\begin{flushright}
\textsf{DESY 09-215}
\end{flushright}

\end{abstract}

\section{Introduction}
An Axion-Like Particle (ALP) is  any
scalar or pseudo-scalar field which couples to the kinetic terms of the
photon.
The pseudo scalar coupling to photons is identical to that of the
axion;  $\mathcal{L} \supset
\frac{\phi}{4M}\epsilon_{\mu\nu\lambda\rho}F^{\mu\nu}F^{\lambda\rho}$,
and a scalar
field couples through the Lagrangian term; $\mathcal{L} \supset
\frac{\phi}{4M}F_{\mu\nu}F^{\mu\nu}$.
The 
presence of contact interactions between ALPs and photons means that
ALPs affect the propagation of photons 
through a magnetic field.  In such an environment a photon can oscillate into an
ALP with probability 
 \cite{Raffelt:1987im}
\begin{equation}
P(z)=\sin^22\theta\sin^2\left(\frac{\Delta(z)}{\cos 2\theta}\right).
\end{equation}
Here $z$ is the distance traveled,  $\Delta(z)=m_{eff}^2z/4\omega$ and $\tan 2\theta
=2B\omega/M m_{eff}^2$. 
$m_{eff}^2=|m_{\phi}^2-\omega_P^2|$,  $m_{\phi}$ is the ALP  mass, $\omega_P$ the plasma frequency of the medium,   $\omega$ the photon frequency
frequency, $B$ the  magnetic field strength and   $M$  the
strength of the photon-ALP coupling.

 In these
proceedings we describe a new test for ALPs  which looks for the
effects  induced by strong ALP-photon mixing on the luminosity of
astronomical objects observed through the magnetic fields of galaxy
clusters.  Our results apply   to  ALPs
with masses $m_{\phi}\lesssim 10^{-12}\mbox{ eV}$. The constraints on
the couplings of such ALPs are:  $10^{11}\mbox{
  GeV}\lesssim M$ for pseudo-scalars \cite{Amsler:2008zzb}, and
$10^{26}\mbox{ GeV} \lesssim M$ for scalars \cite{Will:1993ns}.  However a subclass of scalar
ALPs known as {\em chameleonic} \cite{Khoury:2003rn}
ALPs avoid these constraints  because their mass depends on the local
density, their coupling is required to satisfy  $10^9\mbox{ GeV}\lesssim M$
\cite{Burrage:2008ii}.

\section{Astronomy with ALPs}

The magnetic fields of  galaxy clusters fluctuate on many
different scales. 
 However, at the
high frequencies we consider in what follows   the simple  cell magnetic field
model can be shown to give the same results for ALP-photon mixing as
modeling the variations in the magnetic field with a power spectrum.
The cell model of the magnetic field  assumes the  field is  made up of a large number of equally sized magnetic
domains.  The magnitude of the field strength is the same in each
domain but the orientation of the field is randomly chosen.  
\begin{wrapfigure}{r}{0.45\textwidth}
    \includegraphics[width=80mm]{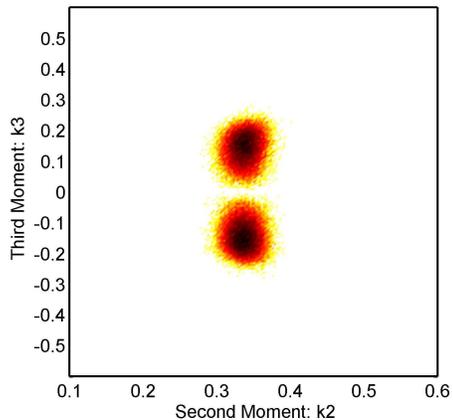}
  \caption{Simulated fingerprint for best fit Gaussian
    model}
 \label{fig:edge-a}
 \end{wrapfigure}

When the probability of mixing between ALPs and photons is
 large the system of photons and ALPs can be evolved
through a large number of  randomly oriented magnetic domains
analytically; this is known as the strong mixing limit.  If $L$ is the size of a magnetic
domain, $N$ the number of domains traversed and $P\equiv P(L)$ is the probability of photon to
ALP conversion in one magnetic domain, we say that strong mixing occurs when $NP \gg 1$
and $N\Delta(L)\lesssim \pi/2$. In this limit the probability of mixing is large, and
frequency independent.
Strong mixing occurs in the magnetic fields of galaxy clusters for
x-ray or  gamma-ray photons  if
$M\lesssim 10^{11}\mbox{ GeV}$, assuming 
  $m_{\phi} \lesssim \omega_P$.\footnote{
  The strength of the magnetic field is $B\approx 1 - 10
\mbox{ $\mu$G}$, the size of a magnetic domain is $L\approx 1\mbox{
  kpc}$ and for a typical source inside the cluster we expect the
light observed from that source to have traversed $N \approx 100 -
1000$ magnetic domains \cite{Carilli:2001hj}.   The plasma frequency in the intracluster
medium is $\omega_P \approx 10^{-12}\mbox{ eV}$} Particles with such masses
and couplings are allowed by current observations for  pseudo-scalar fields and for chameleonic
scalars.

As photon number is not conserved photon-ALP mixing will change the
apparent luminosity of objects observed through the cluster. We  define the
attenuation factor to be the ratio of the flux of photons after
passing through $N$ domains to the initial flux of photons;
$C=I_{\gamma}(N)/I_{\gamma}(0)$. 
Then in the strong mixing limit, assuming no initial flux of ALPs, the mean value for $C$ is $C=2/3$
\cite{Csaki:2001yk}, and its  probability distribution  is  \cite{Burrage:2009mj}
\begin{equation}
f_C(c;p_0)=\frac{1}{\sqrt{1-p_0^2}}\left[\tan^{-1}\left(\sqrt{a}\left(1-\frac{2c_+}{1+p_0}\right)^{-1/2}\right)-\tan^{-1}\left(\sqrt{a}\left(1-\frac{2c_-}{1-p_0}\right)^{1/2}\right)\right],
\label{eq:probdist}
\end{equation}
where $a=(1+p_0)/(1-p_0)$, $c_{\pm}=\mbox{min }(c, (1\pm p_0)/2)$
and $p_0$ is the initial polarization of the photons.
This probability distribution has an unusual shape, and is 
very asymmetric about the mean.  In the next Section  we  show that
this can be exploited as a new test
for ALPs.

\section{Searching for ALPs with luminosity relations}

To use the shape of the probability distribution (\ref{eq:probdist})
to look for ALPs we would need to know the high energy photon flux for
a class of astronomical sources.  We do not currently know of any
objects that are standard candles in x- or gamma-rays,  however, for certain classes of object there exist {\em luminosity
  relations} which correlate the high frequency luminosity of an
object with a feature of its low energy spectrum.
  At low frequencies light mixes weakly with
ALPs and hence  we  assume that low energy observables are not
affected by ALPs at leading order. Therefore luminosity relations can be used to
normalize the high energy flux, so that the effects of ALPs are
observable. \begin{wrapfigure}{r}{0.45\textwidth}
 \includegraphics[width=80mm]{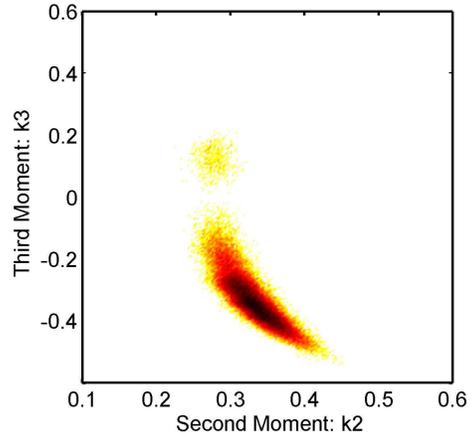}
  \caption{Simulated fingerprint for best fit ALP strong mixing
    model}
 \label{fig:edge-b}
 \end{wrapfigure} The relations typically take the form
\begin{equation}
\log_{10}Y_i=a+b\log_{10}X_i +S_i,
\end{equation}
where  $Y_i$ is the high energy luminosity, and  $X_i$ is the low energy
feature of the spectrum for the $i$-th object in the survey.  $S_i$ represent the scatter in
individual measurements, it is standard in astronomy to assume they
are normally distributed;
$S_i=\sigma\delta_i$ where $\delta \sim N(0,1)$.
If the high frequency light mixes strongly with ALPs this will appear
as an additional contribution to the  scatter 
$S_i=\sigma\delta_i-\log_{10} C_i$, where the $C_i$ are described by
the probability distribution function (\ref{eq:probdist}).

For a given data set we use the likelihood ratio test to see if the
data prefer strong ALP-photon mixing, or the null hypothesis of
Gaussian noise.   We find the values of the parameters $a$, $b$
and $\sigma$ which maximize the likelihood of each hypothesis given the
data, and  then compare these two maximum likelihoods with the ratio $r(p_0)=2\log(\hat{L}_1(p_0)/\hat{L}_0)$,
where $\hat{L}_1(p_0)$ is the maximum likelihood allowing for  strong
ALP-photon mixing  and $\hat{L}_0$ is the
maximum likelihood for models where the scatter  is purely
Gaussian.
 The two hypothesis have the same number
of parameters and therefore $r(p_0)$ is
equivalent to the Bayesian Information Criterion.  
Negative $r(p_0)$ is evidence against ALP strong mixing, and positive
$r(p_0)$ is evidence for ALP strong mixing.  $|r(p_0)|>6$ is
considered strong evidence, $|r(p_0)|>10$ is considered very strong
evidence.

\section{Results from active galactic nuclei}

To apply the test developed in the previous section we require   a
class of compact objects within galaxy clusters  that emit  x-ray or
gamma-ray light and  for which luminosity relations
exist correlating the high energy luminosity with a 
feature of the low-energy spectrum.
 Active galactic nuclei (AGN) satisfy these requirements.  For AGN a luminosity
relation has been established between the 2 keV x-ray luminosity and
the 5 eV optical luminosity.  We have observations of 77 AGN from the
COMBO-17 and ROSAT surveys \cite{Steffen:2006px} and 126 objects from the SDSS
survey \cite{Strateva:2005hu}.  

Applying the likelihood ratio test described in the previous section to
these results we find $r(p_0\lesssim0.5)\approx 25$, where the expectation from AGN physics is that $p_0<0.1$
\cite{1950ratr.book.....C}.

As a qualitative check of this result we plot {\em fingerprints} of the data.
  To do this we construct $10^5$ new data sets, of the same size as
  the original, by bootstrap re-sampling (with replacement) of  the
  original data set.  For each data set we calculate the statistical
  moments of the distribution $k_m({s_i})=[(1/N_p)\sum_is_i^m]^{1/m}$
where  $s_i=\log_{10}Y_i-(a+b\log_{10}X_i)$.  These moments
parametrize the shape
of the probability  distribution.
Fingerprints of the data are then  histogram  plots of  $k_i$ vs.
$k_j$ for the resampled data sets. 
Figures \ref{fig:edge-a} and \ref{fig:edge-b} show example fingerprints for simulated data
respectively without and with the effects of strong ALP-photon
mixing.
\begin{wrapfigure}{r}{0.45\textwidth}
\includegraphics[width=80mm]{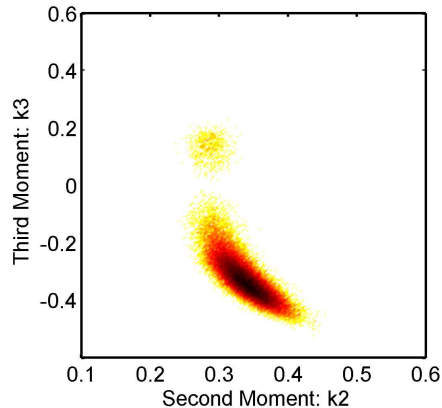}
\caption{Fingerprint from AGN data}
\label{fig1}
\end{wrapfigure}
  Darker regions indicate a higher density of points.  Figure
\ref{fig1} shows the same plot for the data obtained from observations of AGN.

Comparing Figures \ref{fig:edge-b} and \ref{fig1} there is a clear 
qualitative similarity between the shapes of the predicted and
observed 
distributions.   This similarity
persists when higher moments of the distribution are plotted.

The astrophysics underlying the luminosity relation for AGN is not
known, and we cannot rule out that a combination of standard physical
processes in the AGN conspires to mimic the effects of ALP-photon
mixing.  It can be shown, however,  that the scatter in the luminosity
relation is not redshift  dependent, therefore the  observed scatter  is not due to evolution effects or an
incorrect choice of cosmological model.

  \section*{Acknowledgments}

CB acknowledges support from the German Science Foundation (DFG) under 
 	Collaborative Research Centre (SFB) 676. ACD and DJS
        acknowledge STFC.

\begin{footnotesize}

\bibliographystyle{unsrt}
\bibliography{Burrage_Clare.bib}

\end{footnotesize}


\end{document}